\documentclass[11pt]{article}
\usepackage{amsmath}
\usepackage{epsfig}
\usepackage{amsfonts}

\newcommand{\R}{\ensuremath{{\cal R}}}

\newcommand{\del}{\ensuremath{\partial}}

\newcommand{\half}{\ensuremath{\frac{1}{2}}}

\def\Lagrangian{{\cal L}}

\newcommand{\be}{\begin{equation}}
\newcommand{\ee}{\end{equation}}

\newcommand{\ba}{\begin{eqnarray}}
\newcommand{\ea}{\end{eqnarray}}

\newcommand{\ns}{\normalsize}

\begin{document}


\begin{titlepage}

\title{
   \hfill{\ns hep-th/0604046\\}
   \vskip 2cm
   {\Large\bf Scaling solutions and geodesics in moduli space}
\\[0.5cm]}
   \setcounter{footnote}{0}
\author{
{\ns\large 
  \setcounter{footnote}{3}
  J.L.P.Karthauser$^1$\footnote{email: jlk23@sussex.ac.uk}~, 
  P.M.Saffin$^{1,2}$\footnote{email: paul.saffin@nottingham.ac.uk}}
\\[0.5cm]
   $^1${\it\ns Department of Physics and Astronomy, University of Sussex}\\
   {\it\ns Falmer, Brighton BN1 9QJ, UK} \\[0.2em] 
   $^2${\it\ns School of Physics and Astronomy, University of Nottingham}\\
   {\it\ns University Park, Nottingham NG7 2RD, UK}
}

\date{}

\maketitle

\begin{abstract}\noindent
In this paper we consider cosmological scaling solutions in general relativity coupled to
scalar fields with a non-trivial moduli space metric. We discover that the scaling
property of the cosmology is synonymous with the scalar fields tracing out a particular
class of geodesics in moduli space - those which are constructed as integral curves of
the gradient of the log of the potential. 
Given a generic scalar potential we explicitly
construct a moduli metric that allows scaling solutions, and we show the converse - how
one can construct a potential that allows scaling once the moduli metric is known.
\end{abstract}

\thispagestyle{empty}
\end{titlepage}

\section{Introduction}
\label{sec:introduction}


Scaling solutions to the Friedman equation are of prime interest
in modern cosmology, both for study of early time inflationary
scenarios and, more recently, the late time accelerating expansion
of the universe which is suspected to be driven by some dark energy
component in the cosmological
fluid \cite{Padmanabhan:2006ag,Copeland:2006wr}.
The utility of these scaling regimes is that the various components can
evolve such that 
a constant ratio is maintained between their energy densities, and as attractor solutions
they approach these ratios without fine tuning of the initial conditions
and so can be used
to explain the cosmic coincidence of energy densities \cite{Zlatev:1998tr,Wang:1999fa}.
Scaling solutions also find a use in the earlier epoch of inflation, where exponential
potentials \cite{Halliwell:1986ja,Wetterich:1987fm,Burd:1988ss} allow for exact solutions
and sums of exponentials can aid inflationary behaviour \cite{Liddle:1998jc}.
The dynamics of scalar fields with such potentials is now well understood, and
is neatly expressed using ideas from dynamical systems for single scalars
\cite{Copeland:1997et} and multiple scalars 
\cite{Copeland:1999cs,Collinucci:2004iw,Hartong:2006rt}.

That scalar field models can be constructed with potentials that can mimic the
conditions we observe in the universe is but one part of the
story. The theories proposed require a fine-tuning to match
observation, and we are still as of yet to detect the direct signature
of scalar fields in nature; is there any fundamental physics that
might motivate their existence? Fortunately scalar fields abound
in the low energy limit of many modern particle physics models where
they appear dynamically in the form of moduli which parametrize the
size and shape of the geometry of internal dimensions. Interestingly
in many of these models, particularly unified gravity models with
dimensional reduction, gaugino condensation or instanton corrections,
the low energy effective potentials are often exponential in form
\cite{dewit:1963,Biswas:2004pd,Chen:2003dc,Neupane:2005nb,Neupane:2005ms,Karthauser:2006wb,Derendinger:1985kk,Dine:1985rz,Witten:1996bn,Moore:2000fs}.
It would be intriguing indeed if the scalar potentials used
in inflationary theory and quintessence
were in fact a consequence of dynamics of
some extra dimensional string or M-theory.

However, exponential potentials are but one class of possible
potential. How might we investigate the scaling behaviour of more
general potentials? As cosmologists we are well aware of the
geometric nature of pure gravity in Einstein's equations, however
when we couple in additional scalar fields they also come with their
own geometry. As the system evolves one can think of the kinetic term
providing a measure of distance in field space.
This is by virtue
of the existence of a moduli space metric, which for minimal kinetic
terms
is flat, $\partial_\mu \phi^a \partial^\mu \phi^a \equiv \delta_{ab}
\partial_\mu \phi^a \partial^\mu \phi^b $, but there is no reason {\it a priori}
not to also consider other forms for this metric too, perhaps
providing a useful geometric insight into the dynamics of the
system.

There is already a tradition of the study of moduli space metrics
within particle physics models, particularly in supersymmetric field theories
where the field space geometry is usually of some special type, for example
K\"ahler geometry.
More recently
Townsend and Wohlfarth proposed thinking of cosmological evolution
in an augmented field space \cite{Townsend:2004zp},
where the scale factor is added to the geometry of the field space.
It is then discovered that cosmological solutions can be thought of
as geodesics on this extended manifold.
Here however we do not consider this augmented space,
rather we look at the field moduli space alone and discover that in
a scaling regime the fields trace out a particular type of geodesic,
namely a geodesic which coincides with an integral curve of $\del_a\ln(V)$.
Given that the moduli metric, $G_{ab}$, and the potential, $V$, are in principle
independent there may be no such geodesic, or indeed there may be 
many congruences of such geodesics. Here we explicitly 
construct a $G_{ab}$ (although
there may also be others) that allows scaling solutions with a generic potential 
and we also show how one can find a potential
that allows scaling if $G_{ab}$ is known.

\section{Scaling Cosmology and Geodesics}
\label{sec:geodesics}

The starting point for this work is a simple observation following from
previous studies of multi-field systems with $G_{ab}=\delta_{ab}$.
It has been shown \cite{Copeland:1999cs,Collinucci:2004iw,Hartong:2006rt}
that the scaling regime is characterized by
\ba
\dot\phi^a&=&\frac{A^a}{t},\\
\Rightarrow\phi^a&=&A^a\ln(t)+B^a.
\ea
So given two different fields we have
\ba
\phi^a&=&\left(\frac{A^a}{A^b}\right)\phi^b+\left(B^a-\frac{A^a}{A^b}B^b\right),
\ea
which is just the equation of a straight line. As the moduli metric is flat for these cases
then this straight line is, of course, a geodesic. It is our aim to show that even for 
$G_{ab}\neq\delta_{ab}$ this relation between scaling and geodesics persists.

We start by describing the system we aim to study, namely a set of scalar
fields with moduli metric $G_{ab}$ and potential $V$
coupled to gravity. The Lagrangian is given by
\ba
\Lagrangian = \sqrt{-g} \left[ \frac{1}{2\kappa^2} \R
    - \half  G_{ab}\partial_\mu \phi^a\partial^\mu \phi^b - V(\phi) \right],
\ea
with the indices $a,b,c,... \in \{1,2,...,n\}$ running over the
number of scalar fields. We adopt a conventional Friedman-Robertson-Walker
cosmology\footnote{In this paper we take
a space-time metric of signature $(- + + +)$.} for the space-time
metric and restrict the scalar to vary only with time.
This then allows us to write down the effective Lagrangian for
the reduced system, where the dynamical variables depend only on time,
\ba
\Lagrangian = a^3 \left(\half G_{ab} \dot\phi^a \dot\phi^b - V(\phi)\right)
	- \frac{3}{\kappa^2} a \dot{a}^2.
\ea
Here $a = a(t)$ is the scale factor of the universe, and $\,\dot{}\,
\equiv \partial_t = \frac{\partial\phantom{t}}{\partial t}$ denotes
the derivative with respect to time. In this language we find that
the equations of motion, along with the vanishing of the Hamiltonian
due to invariance under time reparametrization, are
\ba
\label{eqn:friedman}
H^2&=&\frac{\kappa^2}{3}\left(\half G_{ab}\dot{\phi}^a\dot{\phi}^b+V\right),\\
\label{eqn:Hdot}
\dot H+3H^2&=&\kappa^2 V,\\
\label{eqn:eomPhi}
\ddot \phi^a+\Gamma^a_{\;bc}\dot\phi^b\dot\phi^c+3H\dot\phi^a&=&-G^{ab}\del_b
V,
\ea
with the Hubble parameter $H \equiv \frac{\dot{a}}{a}$ and the affine connection
$\Gamma^a_{\;bc}$ of $G_{ab}$ defined in the conventional way,
\ba
\Gamma^a_{\;bc} = \half G^{ad} \left( G_{bd,c} + G_{cd,b} - G_{bc,d}
\right).
\ea

Given that we shall be discussing geodesics we now give the standard expressions
for a geodesic curve, ${\cal C}$, with tangent vector $T$ that is parametrized by an affine
parameter $\lambda$.
\ba
T^a&=&\frac{d}{d\lambda}\phi^a,\\
\label{eqn:geodDef}
T^a\nabla_aT^b&=&0\quad\Rightarrow\quad\phi''^a+\Gamma^a_{\;bc}\phi'^b\phi'^c=0,\\
\label{eqn:tanNorm}
T^aT_a&=&1,
\ea
where $'\equiv\frac{d}{d\lambda}$.

We shall take {\it scaling} to mean that the terms in (\ref{eqn:friedman}) 
and (\ref{eqn:Hdot}) evolve in
constant proportion to one another,
$H^2 \propto \dot H \propto V \propto G_{ab} \dot\phi^a \dot\phi^b$.
We start by considering
$H^2$ and the kinetic term of (\ref{eqn:friedman}) and enforce
proportionality by introducing a constant $\alpha$,

\ba
\label{eqn:alphaDef}
\alpha^2 H^2=G_{ab}\dot\phi^a\dot\phi^b.
\ea
We now want to see if this can be consistent with $\phi^a(t)$ tracing out a geodesic,
meaning that the affine parameter can now be thought of as a function of cosmic time
$\lambda=\lambda(t)$. The normalisation of the tangent vector (\ref{eqn:tanNorm})
then gives us
\ba
\label{eqn:kinNorm}
G_{ab}\dot\phi^a\dot\phi^b=\left(\frac{d\lambda}{dt}\right)^2,
\ea
which combines with (\ref{eqn:alphaDef}) to produce
\ba
a=a_0 \exp(\lambda/\alpha).
\ea
Converting time derivatives into $\lambda$ derivatives and using
(\ref{eqn:Hdot}) along with (\ref{eqn:geodDef})
one finds that (\ref{eqn:eomPhi}) becomes
\ba
\label{eqn:phiPrime}
\phi'^a&=&-\frac{1}{\alpha\kappa^2} G^{ab}\del_b \ln(V).
\ea
This is very suggestive, and to make manifest its relation
to geodesics more explicit we shall write it as the components of the tangent co-vector,
\ba
\label{eqn:gradVgeod}
T_a&=& \del_a W, \qquad W = -\frac{1}{\alpha\kappa^2} \ln(V).
\ea
So, if our scaling/geodesic ansatz is consistent with the other terms in the
equations we see that the scaling geodesic takes a very particular form,
it is the integral curve of $\nabla W$. An immediate consequence of this is that one
can never get a scaling solution on closed geodesics.

We also need to check that the potential evolves in constant proportion to $H^2$,
which we can do by using
the fact that these tangent vectors have been normalized
to unity,
\ba
\label{eqn:potOnGeod}
\phi'^a (-\frac{1}{\alpha\kappa^2}) \partial_a \ln(V)=1,
\ea
which is integrated to give
\ba
\label{eqn:Vlambda}
V&=&V_0\exp(-\alpha\kappa^2\lambda)=V_0
\left(\frac{a}{a_0}\right)^{-\alpha^2\kappa^2}.
\label{eqn:constpot}
\ea
If it is the case that $V\propto H^2$ then we can integrate to find
\ba
\label{eqn:scaleFactor}
a^{\alpha^2 \kappa^2 / 2} \propto(t - t_0),
\ea
which, when substituted back into the potential
(\ref{eqn:constpot}), gives the conclusion that
\ba
V \propto \frac{1}{t^2}.
\ea
This is precisely the form that $V$ must take as is easily seen by inspecting
(\ref{eqn:friedman})(\ref{eqn:Hdot}). We also note from (\ref{eqn:scaleFactor})
that we get an accelerating scale factor if $\alpha\kappa<\sqrt 2$.
We can understand such a requirement by reconsidering
(\ref{eqn:tanNorm}) and (\ref{eqn:gradVgeod})
and writing them in the form
\ba
\frac{1}{2\kappa^2}G^{ab}\del_aV\del_bV/V^2=\half\alpha^2\kappa^2,
\ea
and now we recognize the left hand side to be the multi-field generalisation of the slow-roll
parameter $\epsilon$ \cite{Burgess:2004kv}.
We should point out that the potential is only required to take the form
(\ref{eqn:constpot}) on the geodesic itself, away from the geodesic the
potential can take any form, or even be undefined.

A brief comment is in order about the nature of $\alpha$. We see
from (\ref{eqn:friedman}) and (\ref{eqn:alphaDef}) that
\ba
\label{eqn:Tscal}
\tilde{T}&:=& \half \frac{G_{ab} \, \dot\phi^a \dot\phi^b}{H^2} =\half\alpha^2,\\
\label{eqn:Vscal}
\tilde{V}&:=& \frac{V}{H^2} = \frac{3}{\kappa^2} - \half \alpha^2,
\ea
and so it seems in scaling solutions $\alpha$ determines the
value of both the kinetic and potential energies. Moreover, if the potential is
positive then we have an upper bound on $\alpha$ given by $\alpha\kappa<\sqrt 6$.

\section{Construction of a field space metric} \label{sec:metric}

We now come to the issue of whether we can construct a metric $G_{ab}$ that 
will allow scaling for a given potential $V$. Using (\ref{eqn:gradVgeod}) along
with (\ref{eqn:geodDef}) we find that this amounts to solving
\ba
\label{eqn:Wgeod}
G^{ab}\del_aW\left(\del_b\del_cW-\Gamma^d_{\;bc}\del_dW\right)&=&0,
\ea
for the metric. At first sight this looks a difficult task, and a general solution probably
is out of reach. However, it is possible to solve this equation if we restrict ourselves
to a conformally flat metric
\ba
ds^2&=&\exp(2\Omega){d\vec{\phi}}{\,}^2,
\ea
with connection components
\ba
\Gamma^a_{\;bc} &=& \delta^a_b \, \Omega,_c + \, \delta^a_c \, \Omega,_b
- \, \delta_{bc} \, \delta^{ad} \, \Omega,_d.
\ea
This is a natural choice in that (\ref{eqn:Wgeod}) is now an equation for a single
scalar function $\Omega(\phi)$ in terms of the single scalar function $V(\phi)$.
Given that we have already normalized the tangent vector in (\ref{eqn:tanNorm})
this immediately gives us the conformal factor
\ba
\label{eqn:confFactor}
\exp(2\Omega)=\sum_a\del_aW\del_aW
=\frac{1}{\alpha^2\kappa^4}\sum_a\frac{\del_aV\del_aV}{V^2}
\ea
which, as is easily checked, does indeed satisfy (\ref{eqn:Wgeod}).
One consequence of this solution is that if $V$ goes through zero with
non-vanishing gradient then the conformal factor diverges there, pushing
such points to an infinite proper distance in moduli space. Hence 
in a scaling solution $V$ will not change sign.
Another important aspect of this solution is that it allows for a scaling solution through
each point in field space where $\ln(V)$ is well defined; it does not restrict to a single
path as is the case for the usual multi-field scaling solution.

The conformally-flat solution we have presented
is not necessarily the only metric on
field space to give scaling solutions,
however we leave investigation of the construction of more general metrics
for future work.

The other way to proceed is to find a potential that yields a
scaling solution for a given metric. We do not have an explicit solution to this
problem but one certainly exists. Once a moduli metric has been
chosen one finds a suitable, i.e. not closed, geodesic ${\cal C}$. Using the affine parameter
of this geodesic one can reconstruct the value of the potential along ${\cal C}$
using (\ref{eqn:Vlambda}). Simply having $V$ along the geodesic is not enough to
keep $\phi^a$ on the scaling solution, we also need $\del_aV$ which we can find
from (\ref{eqn:gradVgeod}) as we know the geodesic's tangent vector. This is rather
like a race track with banked corners to keep the cars from coming off - the
gradient of $V$ keeps the scalars on the scaling solution.
\section{Constructing scaling solutions}
\label{sec:together}

\begin{figure}[b!]
\center
\epsfig{file=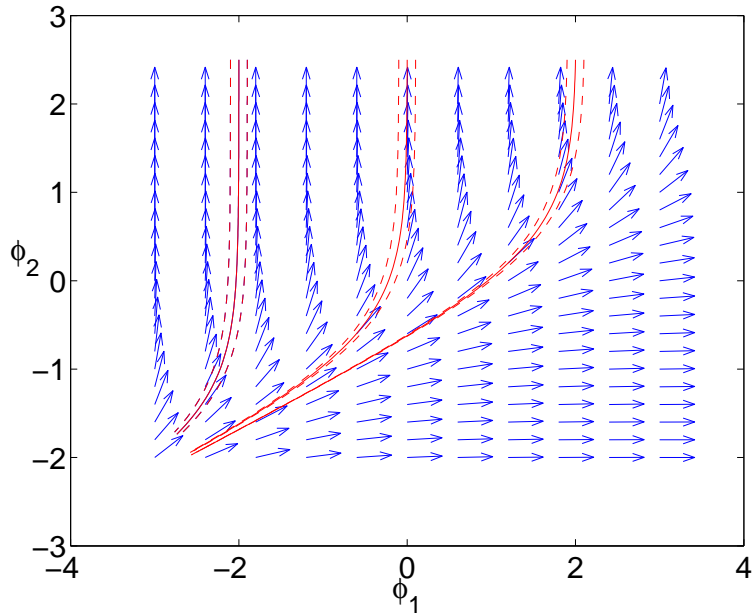,width=10cm}
\flushleft
\caption{
This plot shows the vector field $\nabla \ln V$ of the potential (\ref{testpot1}),
along with some example field evolutions.
The solid lines are the scaling
solutions, with the dashed lines being perturbations about the
scaling solution.
}
\label{fig:grad}
\end{figure}

\begin{figure}
\center
\epsfig{file=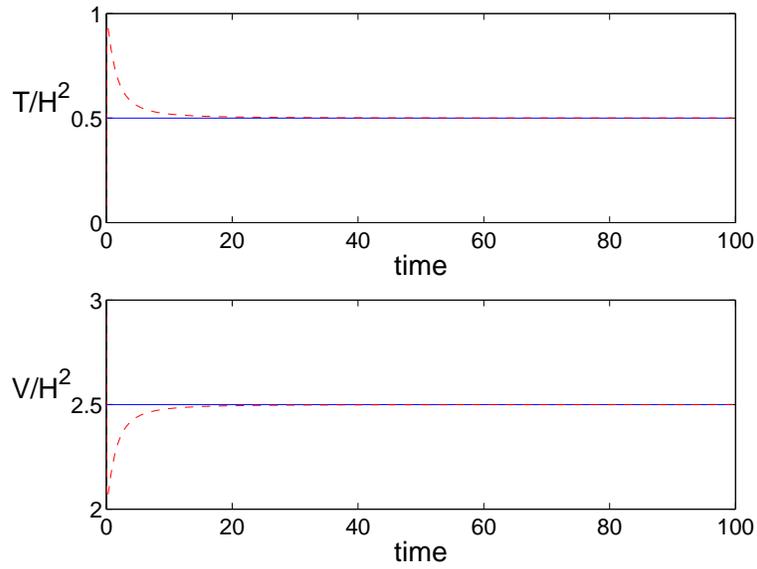,width=10cm}
\flushleft
\caption{
This plot shows the ratio of the kinetic, potential energy to $H^2$
for a scaling solution and perturbations around it.
}
\label{fig:TVscaling}
\end{figure}

\begin{figure}
\center
\epsfig{file=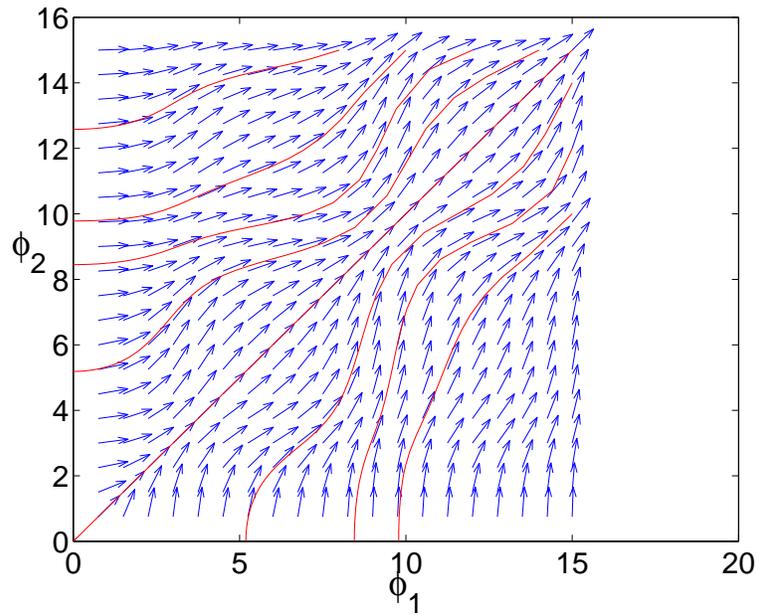,width=10cm}
\flushleft
\caption{
This is the gradient of the potential (\ref{testpot2}), along with
the scaling solutions.
}
\label{fig:grad2}
\end{figure}

Now we have set up the framework for finding scaling solutions in general
potentials we shall give some examples to show how the process works,
by numerically evolving 
(\ref{eqn:friedman})(\ref{eqn:Hdot})(\ref{eqn:eomPhi}).
We shall start with a rather simple example, reminiscent of the multi-field
scaling solutions which required exponential potentials,
\ba
\label{testpot1}
V=\exp(\phi^1)+\exp(2\phi^2),
\ea
using the values $\kappa^2 = 1$ and $\alpha = 1$. In Fig. 
\ref{fig:grad} we have plotted the evolution of $\phi^1$ and $\phi^2$,
also showing the vector field $\del_a \ln V$ (the arrow lengths have been
normalized for clarity).
The solid lines denote scaling solutions and we note
that the fields do indeed follow the gradient of $\ln(V)$ as expected.
The dashed lines represent perturbations of these initial conditions, and
we see that they do not simply go back to the
scaling solutions of the solid lines. This however does not mean that scaling
solutions are not attractors as Fig. \ref{fig:TVscaling} makes clear. Fig. \ref{fig:TVscaling}
is a plot showing the values of the kinetic and potential energies for a
scaling solution (solid line)
and a perturbation of that solution (dashed line). 
What we see is that the perturbed
case does approach a scaling solution, just not the one you started with. This is because
when using the conformally flat moduli metric to solve (\ref{eqn:Wgeod}) there is a scaling
solution through each point in moduli space.
Note also that $T/H^2$, $V/H^2$
approach the values as predicted by (\ref{eqn:Tscal}) and
(\ref{eqn:Vscal}).

As a further test let us consider a more involved, bumpy, potential,
\ba
\label{testpot2}
V&=&(2+\sin(\phi^1))(2+\sin(\phi^2)).
\ea
Scaling solutions for this can be seen in Fig. \ref{fig:grad2}
also following integral curves of the gradient of $\ln(V)$, as our analysis
shows they should.

\section{Adding a barotropic fluid}
\label{sec:fluid}
The effects of adding a baratropic fluid to the system can be studied by
extending our 
Friedman equation
(\ref{eqn:friedman}) and adding the fluid equations,
\ba
H^2&=&\frac{\kappa^2}{3}\left(\half G_{ab}\dot{\phi}^a\dot{\phi}^b+V+\rho\right),\\
P&=&(\gamma-1)\rho,\\
\label{eqn:fluidCont}
\dot\rho&=&-3\gamma H\rho,
\ea
where $P$ is the fluid pressure and $\rho$ is its density.
Such a fluid is typically invoked to represent species of particles
in the early Universe other than the scalar field and so constitutes
an important addition.

Although we have not thoroughly explored the consequences of adding
such a fluid to the system we can find a bound on the equation of state
parameter, $\gamma$, which reduces to the standard one of the usual
scaling analysis.
The analysis follows the same lines as the one presented
above, using the scaling ansatz (\ref{eqn:alphaDef}) as well as 
\ba
\label{eqn:fluidScaling}
V+(1-\gamma/2)\rho&=&\beta V,
\ea
and taking $\beta$ as a new parameter which is constant during scaling.
With this addition we find that (\ref{eqn:phiPrime}) is altered to become
\ba
\phi'^a&=&-\frac{1}{\alpha\beta\kappa^2} G^{ab}\del_b \ln(V).
\ea
A consequence of (\ref{eqn:fluidScaling}) is that if the fluid has $\gamma<2$, which
is physically reasonable, then for positive $V$ and $\rho$ we require $\beta>1$.
This is the origin of the bound we shall derive.

In much the same way that (\ref{eqn:Vlambda}) is derived, one finds that
\ba
\rho\propto a^{-\alpha^2\beta\kappa^2},
\ea
however we also know by integrating (\ref{eqn:fluidCont}) that $\rho\propto a^{-3\gamma}$
which gives us
\ba
\beta&=&3\gamma/(\alpha^2\kappa^2),
\ea
and so our bound on $\beta$ now becomes a bound on the equation of
state parameter
\ba
\label{eqn:bound}
3\gamma>\alpha^2\kappa^2.
\ea
This bound is similar to the one usually found in scaling analyses with a fluid,
indeed taking the case of a single scalar with potential $V=V_0\exp(b\kappa\phi)$
one has that $\alpha^2\kappa^2=9\gamma^2/b^2$, so (\ref{eqn:bound}) becomes
$b^2>3\gamma$ matching the result found in \cite{Copeland:1997et}.

\section{Comments on fixed point analyses}
\label{sec:fixedpoints}

No paper on scaling dynamics appears to be complete without
consideration of the critical point analysis, obtained by recasting
the equations of motion into an autonomous systems
framework \cite{Copeland:1997et,Collinucci:2004iw}.
In this approach the asymptotic behaviour and stability of the system can be
discovered by analysing the nature of fixed point solutions.

The now standard approach for a system with potential
\ba
V&=&\sum\Lambda_i\exp(-\vec\alpha_i.\vec\phi)=\sum V_i
\ea
is to define two new sets of variables
\ba
X^a &\sim&\dot\phi^a/H,\\
Y_i &\sim& V_i/H,
\ea
with which to describe the degrees of freedom.
These are useful because in the case of exponential potentials they are
constant in the scaling regime, giving scaling solutions as fixed points
of the autonomous system.
For the more general case we have been
discussing the closest analogue would be the term
\ba
X\sim \sqrt{G_{ab}\dot\phi^a\dot\phi^b}/H,
\ea
which we know is a constant during scaling. However, this is just one term
and we need to describe $n$ momenta $\dot\phi^a$. In the case where one
uses the conformally flat solution of (\ref{eqn:Wgeod}) one could introduce
\ba
X^a\sim e^\Omega\dot\phi^a/H,
\ea
as a natural set of variables. However, the problem reappears when we try to
find an analogue for the $Y_i$. In general the potential will not split
up nicely into a sum of terms, making the analysis problematic. 
One possible approach to constructing more variables is to use the derivative
of the potential, this is a problem which we hope to return to.

\section{Concluding remarks}
\label{sec:conclusion }

In this paper we studied the scaling solutions of a
Friedman-Robertson-Walker cosmology containing scalar
fields evolving in an arbitrary potential.
The standard picture with canonical kinetic terms would lead
us to conclude that the potential must take a special form, that
of a sum of exponentials, if scaling is to take place.
What we have been able to show is that by including the freedom
of a metric on moduli space we can recover scaling solutions in
generic potentials, with the connection between moduli space geometry
and scalar potential being that the scaling geodesics lie along
integral curves of $\nabla \ln V$.
In fact we have been able to provide a special solution to the
scaling criteria by way of a conformally flat moduli metric, and
we provided explicit examples to show the system in action.
The statement also works in reverse; given a
metric on moduli space it is possible to construct a potential
which will support scaling solutions.
We also briefly considered the effects of adding a baratropic fluid into
the system and were able to find a bound for the equation of state
parameter, with this bound reducing to the standard one when the
moduli metric is flat.

There are a number of things which have not been fully addressed is this
article. Firstly is the issue of stability. While we saw numerical evidence
in our examples that the system approached a scaling solution a more
thorough analysis is required, and we expect it will combine some aspects
of the moduli geometry along with the potential. An approach that could
help with the stability analysis is that of recasting the system in
a form inspired by dynamical systems, writing the equations as an autonomous
system. While we have not presented such a framework we believe
this approach is well worth studying. 

\vspace{1cm}
\noindent
{\large\bf Acknowledgements} The authors would like to
gratefully acknowledge useful discussions with
Mark Hindmarsh and Harvey Reall.
Both JLPK and PMS are supported by PPARC.

\end{document}